\documentstyle[12pt]{article} 
\date{}
\newcommand{\eeq}{\end{eqnarray}}
\newcommand{\beq}{\begin{eqnarray}}

\newcommand{\bB}{{\bf B}}

\newcommand{\vp}{{\varphi}}

\newcommand{\bPi}{\mbox{\boldmath ${\Pi}$}}
\newcommand{\tr}{\mbox{\boldmath $\triangle$}}

\newcommand{\bA}{{\bf A}}

\newcommand{\bx}{{\bf x}}
\newcommand{\by}{{\bf y}}
\newcommand{\bQ}{{\bf Q}}
\newcommand{\nb}{{\nabla}}

\def\con{{}_{\_\rule{-1pt}{0pt}\_}
\rule{-2pt}{0pt}\raise1.5pt\hbox{$\mid$}\hspace{2pt}}
\def\theequation{\arabic{section}.\arabic{equation}}
\newtheorem{TH}{Theorem}
\newtheorem{LM}{Lemma}
\title{Symplectic reduction of $p$-form electrodynamics}
\author{Dariusz Chru\'sci\'nski\footnotemark \\
 Institute of Physics, Nicholas Copernicus University\\
 ul. Grudzi\c{a}dzka 5/7, 87-100 Toru\'n, Poland}

\begin{document}
\def\thefootnote{\relax}\footnotetext{$^*$E-mail:
darch@phys.uni.torun.pl}

\maketitle

\begin{abstract}

We propose a simple method to reduce a general $p$-form electrodynamics with
respect to the standard Gauss constraints. The canonical structure of the
reduced theory  displays a
$p$-dependent sign which makes the essential difference between theories with
different parities of $p$. This feature was observed recently in the
corresponding quantization condition for $p$-brane dyons. It suggests that
these two structures are closely related.

\end{abstract}

\vspace{4cm}

NCU MATHP--99/1

\newpage

\section{Introduction}
\setcounter{equation}{0}

The $p$-form electrodynamics is a simple generalization of an ordinary
electrodynamics in 4-dimensional Minkowski space-time ${\cal M}^4$
where the electromagnetic field potential 1-form $A_\mu$ is replaced by a
$p$-form in $D$-dimensional space-time \cite{Teitel1}, \cite{Teitel2}. The
motivation to study such
type of theories comes e.g. from the string theory where one considers higher
dimensional objects (so called $p$-branes \cite{branes}) interacting with a
gauge field.
   $p$-branes are natural objects which couple in a gauge
invariant way to a $p$-form gauge potential and, therefore, they play a
role of elementary extended sources.

Recently the new input to study a $p$-form theory came
from the electric-magnetic duality which started to play a prominent
role in theoretical physics in the 90. (see e.g. \cite{Olive}).
Now, in order to have
$p$-branes which carry both electric and magnetic charges (i.e. $p$-brane
dyons)  the  dimension of space-time has to be equal $D=2p+2$.
It turns out
\cite{Deser1}, \cite{Gibbons} that there is a crucial difference between
$p$-form theories with different parity of $p$.
Moreover, quantum
mechanics implies the following quantization condition upon the electric and
magnetic charges $(e_1,g_1)$ and $(e_2,g_2)$ of any two dyons:
\beq  \label{q}
e_1g_2 + (-1)^p e_2g_1 = nh\ ,
\eeq
with an integer $n$. For odd $p$ the above condition is a generalization
of the famous Dirac condition \cite{Dirac} but for even $p$ it was
observed only recently \cite{Cremmer}--\cite{Deser3}. Again, a parity of
$p$ plays a crucial role in (\ref{q}).

The aim of the present paper is to show that this $p$-dependence is
already present on the level of a canonical structure of the underlying
$p$-form theory. This feature is however hidden in a standard formulation.
Therefore, we propose a different approach.
Because a theory possesses a rich abelian gauge symmetry we
propose to reduce it with respect to the corresponding Gauss constraints.
Such a reduction
was proposed in the case of ordinary electrodynamics (and linearized
gravity) in
\cite{Jacek}  (it was applied to the variational formulation of the
electrodynamics of point-like charges in \cite{Kij-Dar} and \cite{Darek}).
The method of this paper
 is a straightforward generalization of \cite{Jacek} to a $p$-form case. We
show that the reduced canonical structure (see formula (\ref{Omega-FF}))
displays a $p$-dependent sign in a
manner very similar to (\ref{q}) which shows that duality invariance,
quantization condition and canonical structure are closely related. We
postpone the study of this relation to the next paper. It would be also
interesting to investigate relation between our approach and the standard two
potential formulation of a $p$-form theory (see e.g. \cite{Deser1} and
\cite{SS}).

The  present paper is organized as follows: in section~2 we introduce basic
facts about $p$-form electrodynamics and present its (unreduced) canonical
structure in section~3. Then in section~4 we describe a reduction
procedure. As an example of a reduced $p$-form theory we present a
Maxwell-type theory in section~5 . All technical details and proofs are
contained in appendixes.

\section{$p$-form electrodynamics}
\setcounter{equation}{0}

Consider a $p$-form potential $A$ defined in the $D=2p+2$ dimensional Minkowski
space-time ${\cal M}^{2p+2}$ with the signature of the metric tensor
$(-,+,...,+)$.
The corresponding field tensor is defined as a $(p+1)$-form by $F=dA$:
\beq   \label{F}
F_{\mu_1...\mu_{p+1}} = \partial_{[\mu_1}A_{\mu_2...\mu_{p+1}]}\ ,
\eeq
where the antisymmetrization is taken with a weight one, i.e. $X_{[kl]} :=
X_{kl} - X_{lk}$. Having a Lagrangian
$L$ of the theory one defines another $(p+1)$-form $G$ as follows:
\beq   \label{const}
G^{\mu_1...\mu_{p+1}} = - (p+1)!\frac{\partial L}{\partial
F_{\mu_1...\mu_{p+1}}}\ .
\eeq
Now one may define the electric and magnetic intensities and inductions in the
obvious way:
\beq   \label{E}
E_{i_1...i_p} &=& F_{i_1...i_p0}\ ,\\   \label{B}
B_{i_1...i_p} &=& \frac{1}{(p+1)!}\
\epsilon_{i_1...i_pj_1...j_{p+1}} F^{j_1...j_{p+1}}\ , \\   \label{D}
 D_{i_1...i_p} &=& G_{i_1...i_p0}\ ,\\         \label{H}
H_{i_1...i_p} &=& \frac{1}{(p+1)!} \
\epsilon_{i_1...i_pj_1...j_{p+1}} G^{j_1...j_{p+1}}\ ,
\eeq
where the indices $i_1,i_2,...,j_1,j_2,...$ run from 1 up to $2p+1$ and
$\epsilon_{i_1i_2...i_{2p+1}}$ is the L\'evi-Civita tensor in $2p+1$
dimensional Euclidean space, i.e. a space-like hyperplane $\Sigma$ in the
Minkowski space-time.
The field equations are given by the Bianchi identities $dF=0$, or
in components
\beq   \label{Bianchi}
\partial_{[\lambda}F_{\mu_1...\mu_{p+1}]} =0\ ,
\eeq
and the {\em true} dynamical equations $d*G=0$, or equivalently
\beq  \label{dynamical}
\partial_{[\lambda}*G_{\mu_1...\mu_{p+1}]} =0\ ,
\eeq
 where the Hodge star
operation in ${\cal M}^{2p+2}$ is defined by:
\beq
*G^{\mu_1...\mu_{p+1}} = \frac{1}{(p+1)!} \ \eta^{\mu_1...\mu_{p+1}
 \nu_1...\nu_{p+1}}\ G_{\nu_1...\nu_{p+1}}\
\eeq
and $\eta^{\mu_1\mu_2...\mu_{2p+2}}$ is the covariantly constant volume form
in the Minkowski space-time. Note, that $\epsilon^{i_1...i_{2p+1}} :=
\eta^{0i_1...i_{2p+1}}$.
In terms of electric and magnetic fields defined
in (\ref{E})--(\ref{H}) the field equations (\ref{Bianchi})--(\ref{dynamical})
have the following form:
\beq               \label{dB}
\partial_0 B^{i_1...i_p} &=& (-1)^p \frac{1}{p!}\
\epsilon^{i_1...i_pkj_1...j_p}\
\nb_k E_{j_1...j_p}\ ,\\    \label{B1}
\nb_{i_1} B^{i_1...i_p} &=&0\ ,\\        \label{dD}
\partial_0 D^{i_1...i_p} &=&  \frac{1}{p!}\
\epsilon^{i_1...i_pkj_1...j_p}\
\nb_k H_{j_1...j_p}\ ,\\              \label{D1}
\nb_{i_1} D^{i_1...i_p} &=&0\ ,
\eeq
where $\nb_k$ denotes the covariant derivative on $\Sigma$ compatible with the
metric $\eta_{kl}$ induced from ${\cal M}^{2p+2}$. The
L\'evi-Civita tensor density satisfies
$\epsilon_{12...2p+1} = \sqrt{g}$,
with $g=\det(\eta_{kl})$.

The field equations (\ref{dB})--(\ref{D1}) may be rewritten using
the language of differentials forms. Obviously $D,B,E,H$ are $p$-forms on
$\Sigma$. Let us define $(p+1)$-forms:
\beq
{\cal D} := *_{\Sigma} D\ ,\ \ \ \ \
{\cal B} := *_{\Sigma} B\ ,
\eeq
where ``$*_{\Sigma}$'' denotes the Hodge star on $\Sigma$: for any k-form
\begin{equation}
(*_{\Sigma}X)_{i_1...i_{n-k}} := \frac{1}{(n-k)!}\
\epsilon_{i_1...i_{n-k}j_1...j_k}\ X^{j_1...j_k}\ ,
\end{equation}
with $n=2p+1$.
With this notation one has:
\beq
\dot{{\cal B}} &=& (-1)^p\ dE\ ,\ \ \ \ \ d{\cal B}=0\ ,\\
\dot{{\cal D}} &=&  dH\ ,\ \ \ \ \ \ \ \ \ \ \ \ \ d{\cal D}=0\ .
\eeq

\section{Canonical structure}
\setcounter{equation}{0}

The phase space $\cal P$ of gauge invariant configurations
$({\cal D},{\cal B})$ or equivalently
$(D,B)$ is endowed with the canonical structure.
Using local
coordinates $(D^{i_1...i_p},B^{j_1...j_p})$
 the most general 2-form on $\cal P$  is given  by:
\beq
\Omega_p &=& \int d^{2p+1}x \int d^{2p+1}y \left[
X_{i_1...i_pj_1...j_p}(\bx,\by)
\delta D^{i_1...i_p}(\bx) \wedge \delta B^{j_1...j_p}(\by) \right.\nonumber\\
&+& \frac 12 Y_{i_1...i_pj_1...j_p}(\bx,\by)
\delta D^{i_1...i_p}(\bx) \wedge \delta D^{j_1...j_p}(\by)
 \nonumber\\
&+& \left. \frac 12 Z_{i_1...i_pj_1...j_p}(\bx,\by)
\delta B^{i_1...i_p}(\bx) \wedge \delta B^{j_1...j_p}(\by)
\right] \ .
\eeq
Taking into account the field equations (\ref{dB}) and (\ref{dD}) the
hamiltonian vector field $X_{H_p}$
\beq   \label{ham-rel}
 X_{H_p}\con \Omega_p = - \delta H_p
\eeq
is described by
\beq \label{XH}
X_{H_p} = \int d^{2p+1}x \ \epsilon_{i_1...i_pkj_1...j_p} \left(
\nb^k H^{j_1...j_p} \frac{\delta}{\delta D_{i_1...i_p}}
+ (-1)^p \nb^k E^{j_1...j_p} \frac{\delta}{\delta B_{i_1...i_p}} \right)\ .
\eeq
Therefore, the formula (\ref{ham-rel}) is solved by $H_p$ and $\Omega_p$
defined as follows:
\beq
H_p = \frac{1}{2p!} \int d^{2p+1}x (D^{i_1...i_p}E_{i_1...i_p} +
B^{i_1...i_p}H_{i_1...i_p})\ ,
\eeq
and
\beq    \label{Omega-gi}
\Omega_p =  \int d^{2p+1}x \int d^{2p+1}y\
\delta D^{i_1...i_p}(\bx) \wedge \delta B^{j_1...j_p}(\by)
\epsilon_{i_1...i_pkj_1...j_p} \nb^k \ G(\bx,\by)\ ,
\eeq
where $G(\bx,\by)$ denotes the Coulomb Green function in $2p+1$
dimensional Euclidean space, i.e.
\beq
\triangle \ G(\bx,\by) = - \delta^{(2p+1)}(\bx-\by)\ .
\eeq
The formula (\ref{Omega-gi}) uses only gauge-invariant quantities. It may be,
however, considerably simplified by introducing a gauge potential. One solves
(\ref{B1}) {\em via}:
\begin{equation}  \label{B-A}
B^{i_1...i_p} = \epsilon^{i_1...i_pj_1...j_{p+1}}\, \nb_{j_1}
A_{j_2...j_{p+1}}\ ,
\end{equation}
which is the obvious generalization of $\bB = \nabla \times \bA$ from $p=1$
electrodynamics. Now, modulo gauge-dependent boundary term, the symplectic
form reads:
\begin{equation}
\Omega_p = (-1)^p \int d^{2p+1}x\
\delta D^{i_1...i_p} \wedge \delta A_{i_1...i_p} =
(-1)^{p+1}\int d^{2p+1}x\
\delta G^{\mu_1...\mu_p0} \wedge \delta A_{\mu_1...\mu_p} \ .
\end{equation}
Having a symplectic structure in $\cal P$ one may easily derive corresponding
Poisson brackets:
\begin{equation}   \label{Poisson}
\{ {\cal F},{\cal G}\}_p = \int d^{2p+1}x \left[
\frac{\delta{\cal F}}{\delta D^{i_1...i_p}}\ \epsilon^{i_1...i_pkj_1...j_p}\
\nb_k \frac{\delta{\cal G}}{\delta B^{j_1...j_p}} -
({\cal F} \rightleftharpoons {\cal G}) \right]\ ,
\end{equation}
where $\cal F$ and $\cal G$ are functionals on $\cal P$.
Note, that $D^{i_1...i_p}$ and
$B^{i_1...i_p}$ are not arbitrary but they have to satisfy the Gauss laws
(\ref{B1}) and (\ref{D1}). The fundamental commutation relations implied by
(\ref{Poisson}) read:
\begin{eqnarray}
\{ D^{i_1...i_p}(\bx),B^{j_1...j_p}(\by) \}_p & =&
\epsilon^{i_1...i_pkj_1...j_k}
\nb_k \delta^{(2p+1)}(\bx-\by)\ ,\\
\{ D^{i_1...i_p}(\bx),D^{j_1...j_p}(\by) \}_p &=&
\{ B^{i_1...i_p}(\bx),B^{j_1...j_p}(\by) \}_p  = 0\ ,
\end{eqnarray}
with $\bx,\by\in\Sigma$.

\section{Reduction}
\label{REDUK}
\setcounter{equation}{0}

The idea of this section is to find a new set of electromagnetical variables
using a philosophy of ``$2p+1$" decomposition (it is a straightforward
generalization of \cite{Jacek}). These new variables solve
partially the Gauss constraints (\ref{B1}) and (\ref{D1}). Moreover, they are
canonical with respect to $\Omega_p$.

\subsection{``2p+1" decomposition}

The ``2p+1" decomposition is based on the observation that the $2p+1$
dimensional Euclidean space $\Sigma$ may be decomposed as follows:
\begin{equation}
\Sigma = S^{2p}(1) \times {\bf R}\ ,
\end{equation}
where $S^{2p}(1)$ denotes $2p$ dimensional unit sphere and {\bf R} represents
``radial" direction. Let us introduce spherical coordinates on $\Sigma$:
\beq   \label{s1}
x^A &=& \vp_A\ ,\ \ \ \ \ A=1,2,...,2p\ ,\\    \label{s2}
x^{2p+1} &=& r\ ,
\eeq
where $\vp_1,\vp_2,...,\vp_{2p}$ denote spherical angles (to enumerate angles
we shall use capital letters $A,B,C,...$) (for more details see
Appendix~\ref{SPHERE}). Let $\eta_{kl}$ denote the Minkowskian metric on
$\Sigma$ and let $\lambda_p$ denote the corresponding volume element, i.e.
$\lambda_p = \sqrt{\det\ \eta_{kl}}$
(see (\ref{lambda})).
Finally, let
$\epsilon_{A_1...A_{2p}}$ denote the L\'evi-Civita tensor on $S^{2p}(r)$
such that
$\epsilon_{A_1...A_{2p}} :=\epsilon_{rA_1...A_{2p}}\ $
(we use the same letter to denote the L\'evi-Civita tensors on $\Sigma$ and
$S^{2p}(r)$).
We shall denote by "$|$" the covariant derivative defined by $\eta_{kl}$ on
$\Sigma$ and by "$||$" the covariant derivative on each $S^{2p}(r)$ defined by
the induced metric $\eta_{AB}$.

\subsection{Symplectic reduction}

 Let us consider the evolution
of the $p$-form electromagnetic field in the finite volume $V\in \Sigma$. For
simplicity we take $V=K^{2p+1}(0,R)$ (a $(2p+1)$ dimensional ball of radius
$R$). Of course one may take any $V$ topologically equivalent to
$K^{2p+1}(0,R)$ but then instead of spherical coordinates one chooses
coordinates adapted to $V$.

To reduce $\Omega_p$ in $V$ we use the ``2p+1'' decomposition:
\beq   \label{Omega1}
(-1)^{p+1} \Omega_p &=&  \int_{V} \lambda_p\, \delta G^{i_1...i_p0} \wedge
 \delta A_{i_1...i_p}
= - \int_{V} \lambda_p\, \delta D^{i_1...i_p} \wedge \delta A_{i_1...i_p}
\nonumber\\ &=&
- \int_{V} \lambda_p\, \left( p\,\delta D^{ri_2...i_p}\wedge \delta
A_{ri_2...i_p}
+ \delta D^{A_1...A_p}\wedge \delta A_{A_1...A_p} \right)\ .
\eeq
Using  (\ref{B-A}) one obtains
\begin{equation}
B^{A_1...A_p} = \epsilon^{A_1...A_prB_1...B_p}\ \partial_r
A_{B_1...B_p} + p\, \epsilon^{A_1...A_pB_1rB_2...B_p}\ A_{rB_2...B_p||B_1}\
\end{equation}
and hence
\beq
\epsilon_{A_1...A_pB_1...B_p}\ B^{A_1...A_p||B_1} =
p!\ \delta_{[B_1...B_p]}^{\,C_1...C_p}\ (-1)^p
\left[ \partial_r A_{C_1...C_p} - p\, A_{rC_2...C_p||C_1}
\right]^{||B_1} \ ,
\eeq
where we have used the following identity
\begin{equation}         \label{epsilons}
\epsilon^{i_1...i_Nj_1...j_M} \epsilon_{i_1...i_Nl_1...l_M}
= N!\ \delta^{j_1...j_M}_{[l_1...l_M]} :=
 N!\ \delta^{j_1}_{[l_1}...\delta^{j_M}_{l_M]}\
\end{equation}
with arbitrary $N$ and $M$.

Now, to simplify our consideration, let us choose the following  gauge
conditions for a  $p$-form
$A^{A_1...A_p}$ and $(p-1)$-form $A^{rA_2...A_p}$ on each sphere
$S^{2p}(r)$:
\begin{equation}   \label{gauge}
{A^{A_1...A_p}}_{||A_1} = 0\ ,\ \ \ \ \ \
{A^{rA_2...A_p}}_{||A_2} = 0\ .
\end{equation}
In such a gauge
\begin{equation}  \label{eB}
\epsilon^{A_1...A_pB_1...B_p}\ B_{A_1...A_p||B_1} =
(-1)^{p+1} p\,p!\ r^{-2}\ \tr_{p-1} \,A^{rB_2...B_p}\ ,
\end{equation}
where $\tr_{p-1}$ is defined by:
\begin{equation}  \label{tr-p}
\tr_{p-1} := (p-1)!\left[r^2  \nabla^A\nb_A - (p^2-1)\right] \ .
\end{equation}
This operator has a clear geometrical interpretation.
\begin{LM}  \label{Laplacian}
$r^{-2}\tr_{p-1}$ equals to the Laplace-Beltrami operator for
$(p-1)$-forms $X=(X^{A_2...A_p})$ on $S^{2p}(r)$ satisfying
${X^{A_2...A_p}}_{||A_2}=0$, i.e.
\begin{equation}
r^{-2}\tr_{p-1}X = (d^*d + dd^*)X= d^*dX\ .
\end{equation}
\end{LM}
For proof see Appendix~\ref{PROOFS}. The factor $r^2$ in (\ref{tr-p}) makes
$\tr_{p-1}$ $r$-independent, i.e.
\begin{equation}   \label{r-ind}
[\tr_{p-1}, \partial_r] =0\ .
\end{equation}
\begin{LM}            \label{Kernel}
The operator $\tr_{p-1}$ is invertible.
\end{LM}
For proof see Appendix~\ref{PROOFS}.
Denoting by $\tr_{p-1}^{-1}$ the inverse of $\tr_{p-1}$ we finally obtain:
\begin{equation}   \label{Ar}
A^{rB_2...B_p} = (-1)^{p+1}\, \frac{r^{2}}{p\,p!}\, \tr_{p-1}^{-1} \left(
\epsilon^{A_1...A_pB_1...B_p}\ B_{A_1...A_p||B_1} \right) \ .
\end{equation}
Introducing variables:
\beq  \label{Q1}
Q_{(1)}^{A_2...A_p} &:=& r\, D^{rA_2...A_p}\ ,\\   \label{Pi1}
\Pi^{(1)}_{B_2...B_p} &:=&  \frac{r}{p!} \, \tr_{p-1}^{-1} \left(
\epsilon_{A_1...A_pB_1...B_p}\ B^{A_1...A_p||B_1} \right) \
\eeq
we may rewrite the first integral in (\ref{Omega1}) as
\begin{equation}            \label{part1}
- \int_{V} \lambda_p\  p\,\delta D^{ri_2...i_p}\wedge \delta A_{ri_2...i_p}
= (-1)^{p+1} \int_{V} \lambda_p\ \delta \Pi^{(1)}_{A_2...A_p} \wedge
\delta Q_{(1)}^{A_2...A_p}\ .
\end{equation}
To reduce the second term in (\ref{Omega1}) observe that in the special gauge
(\ref{gauge}), $A_{A_1...A_p}$ may be expressed as follows:
\begin{equation}  \label{A-beta}
A_{A_1...A_p} = \epsilon_{A_1...A_pB_1...B_p} \ \beta^{B_2...B_p||B_1}\ .
\end{equation}
Now, from (\ref{B-A})
\begin{equation}        \label{BB-A}
B^{rB_2...B_p} = \epsilon^{B_2...B_pC_1...C_{p+1}}\ A_{C_2...C_{p+1}||C_1}\ .
\end{equation}
Inserting (\ref{A-beta}) into (\ref{BB-A}) and using once more the gauge
conditions (\ref{gauge}) we obtain
\begin{equation}  \label{Br}
B^{rB_2...B_p} = - p!\, r^{-2}\ \tr_{p-1}\, \beta^{B_2...B_p}\ ,
\end{equation}
and, therefore
\begin{equation}
\beta^{B_2...B_p} = - \frac{r^{2}}{p!}\ \tr_{p-1}^{-1}\, B^{rB_2...B_p}\ ,
\end{equation}
which, together with (\ref{A-beta}), gives
\begin{equation}   \label{A}
A_{A_1...A_p} = - \frac{r^{2}}{p!}\ \epsilon_{A_1...A_pB_1...B_p} \
\left( \tr_{p-1}^{-1} \, B^{rB_2...B_p} \right)^{||B_1}\ .
\end{equation}
Having $A_{A_1...A_p}$ we may calculate the second term in (\ref{Omega1}):
\beq  \label{o1}
\lefteqn{
- \int_{V} \lambda_p\  \delta D^{A_1...A_p}\wedge \delta A_{A_1...A_p}
}\nonumber\\
 & & = - \int_{V} \lambda_p\, \frac{r}{p!} \left(
\epsilon_{A_1...A_pB_1...B_p}\
\delta D^{A_1...A_p} \right)^{||B_1} \wedge\ r\, \tr_{p-1}^{-1}\,
\delta B^{rB_2...B_p}\ .
\eeq
Defining:
\beq  \label{Q2}
Q_{(2)}^{A_2...A_p} &:=& r\, B^{rA_2...A_p}\ ,\\   \label{Pi2}
\Pi^{(2)}_{B_2...B_p} &:=& -\frac{r}{p!} \, \tr_{p-1}^{-1} \left(
\epsilon_{A_1...A_pB_1...B_p}\ D^{A_1...A_p||B_1} \right) \
\eeq
we may rewrite (\ref{o1}) as
\begin{equation}   \label{part2}
- \int_{V} \lambda_p\  \delta D^{A_1...A_p}\wedge \delta A_{A_1...A_p}
= \int_{V} \lambda_p\ \delta \Pi^{(2)}_{A_2...A_p} \wedge
\delta Q_{(2)}^{A_2...A_p}\ .
\end{equation}
Now, taking into account (\ref{part1}) and (\ref{part2}) we have finally
\begin{equation}  \label{Omega-F}
\Omega^{red}_p  = \int_{V} \lambda_p \left(
 \delta \Pi^{(1)}_{A_2...A_p} \wedge
\delta Q_{(1)}^{A_2...A_p} + (-1)^{p+1}
\delta \Pi^{(2)}_{A_2...A_p} \wedge
\delta Q_{(2)}^{A_2...A_p} \right) \ .
\end{equation}
Let us use  more compact  notation
\begin{equation}
{\bf X}\cdot{\bf Y} := X^{A_2...A_p}Y_{A_2...A_p}\ .
\end{equation}
Our result may be formulated as
\begin{TH}
The reduced symplectic structure $\Omega^{red}_p$ has on the reduced phase
space ${\cal P}^{red}_p := (\bQ_{(\alpha)},\bPi^{(\alpha)})$ the following
form:
\begin{equation}  \label{Omega-FF}
\Omega^{red}_p  = \int_{V} \lambda_p \left(
 \delta \bPi^{(1)}\cdot \wedge
\delta \bQ_{(1)}
+ (-1)^{p+1} \delta \bPi^{(2)}\cdot \wedge
\delta \bQ_{(2)} \right)\ .
\end{equation}
\end{TH}
The above theorem obviously implies
\begin{TH}
The corresponding reduced Poisson bracket (\ref{Poisson}) reads:
\begin{equation}   \label{Poisson-new}
\{ {\cal F},{\cal G}\}_p^{red} = \int_{V} \lambda_p\, \left[
\frac{\delta{\cal F}}{\delta \bQ_{(1)}} \cdot
\frac{\delta{\cal G}}{\delta \bPi^{(1)}}
+ (-1)^{p+1} \frac{\delta{\cal F}}{\delta \bQ_{(2)}} \cdot
\frac{\delta{\cal G}}{\delta \bPi^{(2)}}
- ({\cal F} \rightleftharpoons {\cal G}) \right]\ .
\end{equation}
\end{TH}
This formula leads to the following commutation relations:
\beq
\{ Q^{A_2...A_p}_{(1)}(\bx),\Pi_{B_2...B_p}^{(1)}(\by)\}_p^{red}
&=&  \delta^{A_2...A_p}_{[B_2...B_p]}\
\delta^{(2p+1)}(\bx-\by)\ , \\
\{ Q^{A_2...A_p}_{(2)}(\bx),\Pi_{B_2...B_p}^{(2)}(\by)\}_p^{red}
&=& (-1)^{p+1} \delta^{A_2...A_p}_{[B_2...B_p]}\
\delta^{(2p+1)}(\bx-\by)\ ,
\eeq
and the remaining brackets vanish.

\subsection{Reconstruction}

Now, could we reconstruct $D$ and $B$ in terms of $\bQ$'s and $\bPi$'s? The
answer is positive by virtue of the following reconstruction
\begin{TH}
The variables $(\bQ_{(\alpha)},\bPi^{(\alpha)};\ \alpha=1,2)$
contain
the entire gauge-invariant information about the fields $D^{i_1...i_p}$ and
$B^{i_1...i_p}$.
\end{TH}
{\em Proof}: we show that $\bQ_{(1)}$ and $\bPi^{(2)}$ encode the information
about $D$ field. Due to (\ref{Q1}) it is enough to show  how to
reconstruct the tangential part of $D$, i.e. $D^{A_1...A_p}$. Let us
make the following decomposition (see Appendix~\ref{SPHERE}):
\begin{equation}
D^{A_1...A_p} = {\cal U}^{[A_2...A_p||A_1]} + \epsilon^{A_1...A_pB_1...B_p}
{\cal V}_{B_2...B_p||B_1}\ .
\end{equation}
\begin{LM}   \label{Dec}
The following identities hold:
\beq     \label{ident1}
r^{2} {D^{A_1...A_p}}_{||A_1} &=& \tr_{p-1} \, {\cal U}^{A_2...A_p}\ ,\\
\label{ident2}
r^{2} \epsilon_{A_1...A_pB_1...B_p}\ D^{A_1...A_p||B_1} &=&
\tr_{p-1}\, {\cal V}_{B_2...B_p}\ .
\eeq
\end{LM}
For proof see Appendix~\ref{PROOFS}.
Therefore
\begin{equation}
{\cal V}_{B_2...B_p} = - p!\, r\, \Pi^{(2)}_{B_2...B_p}\ .
\end{equation}
Now, to reconstruct ${\cal U}^{A_2...A_p}$ let us use (\ref{D1}).
\begin{LM}   \label{GAUSS}
The Gauss constraints ${D^{kA_2...A_p}}_{|k}$   are equivalent to
\begin{equation}   \label{Gauss}
\partial_r (r^{2p}\,D^{rA_2...A_p}) + r^{2p} {D^{A_1A_2...A_p}}_{||A_1}
= 0   \ .
\end{equation}
\end{LM}
The proof see Appendix~\ref{PROOFS}.
Finally, due to (\ref{Gauss})
\begin{equation}
{\cal U}^{A_2...A_p} = - r^2 \tr_{p-1}^{-1}\, \partial_r \left( r^{2p-1}\,
Q_{(1)}^{A_2...A_p}
\right)\ .
\end{equation}
In the same way one shows that $\bQ_{(2)}$ and $\bPi^{(1)}$ reconstruct $B$
field.
\hspace*{\fill}$\Box$

\subsection{Counting degrees of freedom}

Let us count the total number $N_p$ of degrees of freedom for a $p$-form
theory.
Define a number $n^p_k$ by:
\begin{equation}
n^p_k := \left(\begin{array}{c}
       2p+1\\ k \end{array}\right) \ .
\end{equation}
A $p$-form potential $A_{i_1...i_p}$ on $\Sigma$ has $n^p_p$
independent components. However, due to a gauge
freedom
\begin{equation}                   \label{L1}
A_{i_1...i_p} \rightarrow A_{i_1...i_p} +
\nabla_{[i_1}\Lambda^{(1)}_{i_2...i_p]}
\end{equation}
some of them may be gauged away e.g. to zero. $\Lambda^{(1)}_{i_1...i_{p-1}}$
carries $n^p_{p-1}$ components but  the equation (\ref{L1}) is still invariant
under
\begin{equation}    \label{L2}
\Lambda^{(1)}_{i_1...i_{p-1}} \rightarrow \Lambda^{(1)}_{i_1...i_{p-1}} +
\nabla_{[i_1}\Lambda^{(2)}_{i_2...i_{p-1}]}\ .
\end{equation}
Therefore, some of the components of $\Lambda^{(1)}$ may be gauged away {\em
via} $\Lambda^{(2)}$ which has $n^p_{p-2}$ components. The same argument
applies for
$\Lambda^{(2)}$ and one continues up to a 0-form $\Lambda^p$ carrying
obviously only one component.
 Therefore, the total number of degrees of freedom carried by
$A_{i_1...i_p}$ equals:
\begin{equation}
N_p = n^p_p - n^p_{p-1} + n^p_{p-2} - ... + (-1)^p n^p_0\ .
\end{equation}
Using the standard mathematical induction it is easy to prove that
\begin{equation}
N_p =
 \left(\begin{array}{c}
2p\\ p \end{array}\right) \ .
\end{equation}
Now, note that $\bQ_{(1)}$ and $\bQ_{(2)}$ carry together
\begin{equation}
M_p =  2\left(\begin{array}{c}
2p\\ p-1 \end{array}\right) = \frac{2p}{p+1}\, N_p \ .
\end{equation}
Therefore,
\begin{equation}
M_p \geq N_p
\end{equation}
and the equality holds only for $p=1$. It means that for $p>1$,
$\bQ_{(\alpha)}$ are not independent.
\begin{LM}  \label{Q-A}
For $p>1$, reduced variables $\bQ_{(\alpha)}$ and  $\bPi^{(\alpha)}$
 satisfy the following constraints:
\begin{equation}
\nabla_{A_2}{Q_{(\alpha)}^{A_2...A_p}} =0\ , \ \ \ \
\nabla^{A_2}{\Pi^{(\alpha)}_{A_2...A_p}} =0\ .
\end{equation}
\end{LM}
The proof follows obviously from (\ref{B1}) and (\ref{D1}). Note, that these
constraints are trivially satisfied  for $p=1$.

\section{Maxwell theory}
\label{MAXWELL}
\setcounter{equation}{0}

The simplest $p$-form theory is of the Maxwell type, i.e. constitutive relations
between inductions $(D,B)$ and intensities $(E,H)$ are linear:
\begin{equation}
D^{i_1...i_p} = E^{i_1...i_p}\ , \ \ \ \ \
B^{i_1...i_p} = H^{i_1...i_p}\ .
\end{equation}
In such a case it is easy to rewrite field equations (\ref{dB}) and (\ref{dD})
in terms of $\bQ$'s and $\bPi$'s.

\begin{TH}
Equations (\ref{dB}) and (\ref{dD}) are equivalent to:
\beq                       \label{dQ1}
\dot{{Q}}_{(1)}^{A_2...A_p} &=& - \tr_{p-1}\
\Pi_{(1)}^{A_2...A_p}\ ,\\
\label{dQ2}
\dot{{Q}}_{(2)}^{A_2...A_p} &=& (-1)^p \tr_{p-1}\
\Pi_{(2)}^{A_2...A_p}\ ,\\
\label{dP1}
\dot{\Pi}^{(1)}_{A_2...A_p} &=& - \tr_{p-1}^{-1}\left[ r^{-1}
\partial^2_r(rQ^{(1)}_{A_2...A_p}) + r^{-2}
\tr_{p-1}\,Q^{(1)}_{A_2...A_p}
\right]\ ,\\
\label{dP2}
\dot{\Pi}^{(2)}_{A_2...A_p} &=& (-1)^{p} \tr_{p-1}^{-1}\left[ r^{-1}
\partial^2_r(rQ^{(2)}_{A_2...A_p}) + r^{-2}
\tr_{p-1}\,Q^{(2)}_{A_2...A_p}
\right]\ ,
\eeq
where $Q^{(\alpha)}_{A_2...A_p} := \eta_{A_2B_2}...\eta_{A_pB_p}
Q_{(\alpha)}^{B_2...B_p}\ $ and
$\ \Pi_{(\alpha)}^{A_2...A_p} := \eta^{A_2B_2}...\eta^{A_pB_p}
\Pi^{(\alpha)}_{B_2...B_p}\ $.
\end{TH}
For proof see Appendix~\ref{MAX}.
These equations may be derived from the following hamiltonian:
\beq              \label{H-Maxwell}
H_p &=& \frac{1}{2(p-1)!} \int \lambda_p \left[ r^{-2} \bQ_{(\alpha)}
\cdot\bQ^{(\alpha)} - r^{-2p}\partial_r(r^{2p-1}\bQ_{(\alpha)})
\tr_{p-1}^{-1}\cdot \partial_r(r\bQ^{(\alpha)})
\right. \nonumber\\
&-& \left. \bPi_{(\alpha)} \tr_{p-1}\cdot \bPi^{(\alpha)} \right]\ ,
\eeq
i.e.
\beq
\dot{\bQ}_{(\alpha)} &=& \{ {\bQ}_{(\alpha)}, H_p\}_p^{red}\ ,\\
\dot{\bPi}_{(\alpha)} &=& \{ {\bPi}_{(\alpha)}, H_p\}_p^{red}\ ,
\eeq
with the Poisson bracket defined in (\ref{Poisson-new}).
One easily shows that the numerical value
of $H_p$ is equal to the standard (i.e. obtained {\em via} the symmetric
energy-momentum tensor)
electromagnetic energy contained in $V$, i.e.
\begin{equation}
H_p = \frac{1}{2p!} \int_{V} \lambda_p \left( D^{i_1...i_p}D_{i_1...i_p} +
B^{i_1...i_p}B_{i_1...i_p} \right)\ .
\end{equation}

\section*{Appendixes}

\appendix
\def\theequation{\thesection.\arabic{equation}}

\section{Geometry of $S^{2p}(r)$}
\label{SPHERE}
\setcounter{equation}{0}

The
relation between cartesian $(y^1,y^2,...,y^{2p+1})$ and spherical coordinates
$(x^1,x^2,...,x^{2p+1})$ introduced in (\ref{s1})--(\ref{s2})
reads as follows:
\begin{eqnarray*}
y^1 &=& r\sin\vp_{2p}\sin\vp_{2p-1}...\sin\vp_3\sin\vp_2\sin\vp_1\ ,\\
y^2 &=& r\sin\vp_{2p}\sin\vp_{2p-1}...\sin\vp_3\sin\vp_2\cos\vp_1\ ,\\
y^3 &=& r\sin\vp_{2p}\sin\vp_{2p-1}...\sin\vp_3\cos\vp_2\ ,\\
\vdots & & \\
y^{2p} &=& r\sin\vp_{2p}\cos\vp_{2p-1}\ ,\\
y^{2p+1} &=& r\cos\vp_{2p}\ .
\end{eqnarray*}
The Minkowskian metric on $\Sigma$ is diagonal:
\beq
\eta_{kk} &=& r^2 \prod_{l=k+1}^{2p} \sin^2\vp_l\ ,\ \ \ \ \ k=1,2,...,2p-1,
\\
\eta_{2p,2p} &=& r^2\ ,\ \ \ \ \ \eta_{rr} = 1\ .
\eeq
Therefore, the volume element
\begin{equation}  \label{lambda}
\lambda_p = \sqrt{\det\eta_{kl}} = r^{2p}\prod_{A=1}^{2p} \sin^{A-1}\vp_A\ .
\end{equation}
One easily calculates the corresponding Christoffel symbols. The only
nonvanishing symbols are:
\begin{equation}   \label{Gammas}
\Gamma^A_{BC}\ ,\ \ \ \Gamma^A_{Br} = r^{-1}\delta^A_{\ B}\ ,\ \ \
\Gamma^r_{AB} = - r^{-1}\eta_{AB}\ .
\end{equation}
Now, one easily finds the Riemann tensor on $S^{2p}(r)$:
\begin{equation}  \label{Riemann}
R_{ABCD} = r^{-2} (\eta_{AC}\eta_{DB} - \eta_{AD}\eta_{CB})\ ,
\end{equation}
where $\eta_{AB}$ denotes the induced metric on each $S^{2p}(r)$. The
corresponding Ricci tensor $R_{AB}$ and a scalar curvature $R$ read:
\begin{equation}     \label{Ricci}
R_{AB} = \frac{2p-1}{r^2}\ \eta_{AB}\ ,\ \ \ \  R= \frac{2p(2p-1)}{r^2}\ .
\end{equation}
Finally, let
$\epsilon_{A_1...A_{2p}}$ denote the L\'evi-Civita tensor density on
$S^{2p}(r)$ such that
$$
\epsilon_{12...2p} = \lambda_p\ .
$$
Let $X=X_{A_1...A_p}$ denote any $p$-form field on $S^{2p}(r)$. This field may
be decomposed as follows:
\begin{equation}                   \label{dec}
X_{A_1...A_p} = \nabla_{[A_1}\alpha_{A_2...A_p]} +
\epsilon_{A_1...A_pB_1...B_p} \nabla^{B_1}
\beta^{B_2...B_p}\ ,
\end{equation}
Formula (\ref{dec}) follows from the Hodge
theorem
\begin{equation}   \label{Hodge}
X = d\alpha + d^*\beta' + h\ ,
\end{equation}
with $\alpha$ being a $(p-1)$-form, $\beta'$ a $(p+1)$-form and $h$ a harmonic
$p$-form. Note, that there is no harmonic term in (\ref{dec}) because the set of
harmonic $p$-forms on $S^{2p}(r)$ is empty. Obviously, $\beta'=\lambda*\beta$
for some constant $\lambda$. To find $\lambda$ note, that
for even dimensional Riemannian manifolds the adjoint $d^*$ of $d$ reads:
\begin{equation}   \label{d*}
d^* = - *d*\
\end{equation}
and,therefore,
$d^*\beta' = -\lambda *d**\beta$.
Now, on a Riemannian $n$-dimensional manifold
\begin{equation}
**\,\omega_k = (-1)^{k(n-k)}\omega_k\ ,
\end{equation}
and hence for $n=2p$,
$**\beta = (-1)^{p+1} \beta$.
Therefore,
$d^*\beta' = (-1)^p\lambda *d\beta$ which implies $\lambda = (-1)^p$.

\section{Proofs}
\label{PROOFS}
\setcounter{equation}{0}

\subsection{Proof of Lemma~\protect\ref{Laplacian}}

Consider any $(p-1)$-form $X=(X_{A_2...A_p})$ on $S^{2p}(r)$ such that
$\nabla^{A_2}X_{A_2...A_p}=0$. This condition may be rewritten as
\begin{equation}
d^*\,X=0\ .
\end{equation}
Therefore, the Laplacian on $X$ reduces to
\begin{equation}
(d^*d + dd^*)X = d^*d\,X\ .
\end{equation}
Due to (\ref{d*}), $d^*d\,X$ may be calculated as follows:
\beq
(*d*dX)_{A_2...A_p} &=& \frac{1}{(p+1)!} \epsilon_{A_2...A_pAB_1...B_p}
\nabla^{[A}(*dX)^{B_1...B_p]} \nonumber\\
&=& \frac{1}{(p+1)!} \epsilon_{A_2...A_pAB_1...B_p}
\nabla^{[A} \frac{1}{p!} \epsilon^{B_1...B_p]C_1...C_p}
\nabla_{[C_1}X_{C_2...C_p]} \nonumber\\
&=& \frac{p!}{p!(p+1)!} \epsilon_{A_2...A_pAB_1...B_p} \left(
\nabla^A \epsilon^{B_1...B_pC_1...C_p} -
\nabla^{B_1} \epsilon^{AB_2...B_pC_1...C_p}\right.\nonumber\\
 & & \left. -  ... -
\nabla^{B_p} \epsilon^{B_1...B_{p-1}AC_1...C_p} \right)
\nabla_{[C_1}X_{C_2...C_p]} \nonumber\\
&=& \frac{1}{(p+1)!} \epsilon_{A_2...A_pAB_1...B_p} \left(
\epsilon^{B_1...B_pC_1...C_p} \nabla^A -
p\, \epsilon^{AB_2...B_pC_1...C_p} \nabla^{B_1}\right) \nonumber\\
& & \cdot \nabla_{[C_1}X_{C_2...C_p]} \nonumber\\
&=& \frac{(-1)^pp!}{(p+1)!} \left(
\delta^{\ C_1\ ...\ C_p}_{[A_2...A_pA]}\, \nabla^A
+  p \delta^{\ C_1\ ...\ C_p}_{[A_2...A_pB_1]}\, \nabla^{B_1}
\right) \nabla_{[C_1}X_{C_2...C_p]} \nonumber\\
&=& - \delta^{C_1C_2...C_p}_{[AA_2...A_p]}\, \nabla^A
\nabla_{[C_1}X_{C_2...C_p]} \nonumber\\
&=& (p-1)! \left(- \nabla^A\nabla_A\, X_{A_2...A_p} + (-1)^{p}
\nabla^A\nabla_{[A_2}X_{A_3...A_p]A}  \right)\ .
\eeq
To calculate the second term note that
\beq \label{2}
\nabla^A\nabla_{[A_2}X_{A_3...A_p]A} &=&
\nabla^A \left( \nabla_{A_2}X_{A_3...A_pA} - \nabla_{A_3}X_{A_2A_4...A_pA} -
\right. \nonumber\\  &-&  \left. ...   -
\nabla_{A_p} X_{A_2...A_{p-1}A_2A} \right)\ .
\eeq
Now, we use the following well known
property of
the Riemann tensor
\beq  \label{property}
[\nabla_A,\nabla_B] T^{C_1...C_k} =
R^{C_1}_{\ \    DAB}T^{DC_2...C_k} + ... + R^{C_k}_{\ \ DAB}
 T^{C_2...C_{k-1}D} \ ,
\eeq
which holds for any tensor field $T^{C_1...C_k}$. Using (\ref{property}) one
has:
\beq   \label{3}
\nabla_A\nabla^{A_2}X^{A_3...A_pA} &=&
\eta^{A_2B}[\nabla_A,\nabla_B]X^{A_3...A_pA} =
\eta^{A_2B} \left( R^{A_3}_{ \ \ CAB}X^{CA_4...A_pA} \right.\nonumber\\
&  & \left. + ... +
R^{A_p}_{\ \ CAB}X^{A_3...A_{p-1}CA} + R^{A}_{\ \
CAB}X^{A_3...A_{p}C}\right) \ .
\eeq
Using (\ref{Riemann}) one obtains
\beq
\eta^{A_2B}  R^{A_3}_{\ \ \ CAB}X^{CA_4...A_pA} &=&
r^{-2} \eta^{A_2B}\eta^{A_3D}(\eta_{DA}\eta_{CB} - \eta_{DB}\eta_{CA} )
X^{CA_4...A_pA}\nonumber\\
&=& r^{-2}X^{A_2A_4...A_pA_3} =
r^{-2} (-1)^{p+1}X^{A_2...A_p}\ .
\eeq
Exactly the same result holds for the first $(p-2)$ terms in (\ref{3}). The
last term
\beq
\eta^{A_2B}  R^{A}_{\ \ CAB}X^{A_3...A_pC} =
\eta^{A_2B} R_{CB} X^{A_3...A_pC} =
 r^{-2} (-1)^p (2p-1) X^{A_2...A_p}\ ,
\eeq
due to (\ref{Ricci}). Finally,
\beq
(-*d*dX)_{A_2...A_p} &=& (p-1)!\left[\nabla^A\nabla_A - r^{-2}(p^2-1)
\right]X_{A_2...A_p} \nonumber\\ &=&  r^{-2} \tr_{p-1}X_{A_2...A_p}\ .
\eeq
\hspace*{\fill}$\Box$

\subsection{Proof of Lemma~\protect\ref{Kernel}}

It is a well known  fact from the theory of cohomology that on
$S^{n}(r)$ the kernel of $\tr_k$ is nontrivial only for $k=0$ and $k=n$.
 The $p=1$ case (i.e. an ordinary
electrodynamics in ${\cal M}^4$) is special because $\tr_0 = r^2\nb^A\nb_A$
possesses zero modes which are simply constant functions on $S^2(r)$. This
problem was treated in \cite{Jacek}.

\hspace*{\fill}$\Box$

\subsection{Proof of Lemma~\protect\ref{Dec}}

Let $X=(X_{A_1...A_p})$ be a $p$-form on $S^{2p}(r)$.
 Due to the Hodge theorem
(\ref{Hodge}) it may be decomposed as follows:
\begin{equation}
X_{A_1...A_p} = \nabla_{[A_1}\alpha_{A_2...A_p]} +
\epsilon_{A_1...A_pB_1...B_p}\nabla^{B_1}\beta^{B_2...B_p}\ ,
\end{equation}
i.e. $\alpha$ and $\beta$ are $(p-1)$-forms on $S^{2p}(r)$:
\begin{equation}  \label{X}
X = d\alpha + *d\beta\ .
\end{equation}
Let us choose a following ``gauge condition" for $\alpha$ and $\beta$:
\begin{equation}
d^*\alpha = d^*\beta = 0\ .
\end{equation}
Now, taking into account (\ref{X}) and Lemma~\ref{Laplacian} one obtains
\begin{equation}  \label{alpha}
d^*X= d^*d\alpha = r^{-2}\tr_{p-1}\alpha
\end{equation}
and
\begin{equation}    \label{beta}
*dX =  *d*d\beta = - d^*d\beta =
-\tr_{p-1}\beta\ .
\end{equation}
Rewriting (\ref{alpha}) and (\ref{beta}) in components on obtains:
\begin{equation}
r^2 {X^{A_1A_2...A_p}}_{||A_1} = \tr_{p-1}\alpha^{A_2...A_p}\ ,
\end{equation}
and
\beq
\tr_{p-1}\beta^{B_2...B_p} &=&  - r^2  (*dX)^{B_2...B_p} =
\frac{-1}{(p+1)!}
\epsilon^{B_2...B_pB_1A_1...A_p}\nabla_{[B_1}X_{A_1...A_p]}
\nonumber\\
&=&  \epsilon^{A_1...A_pB_1...B_p}\nabla_{B_1}X_{A_1...A_p}\ .
\eeq

\hspace*{\fill}$\Box$

\subsection{Proof of Lemma~\protect\ref{GAUSS}}

\beq
{D^{ki_2...i_p}}_{|k} = \partial_k D^{ki_2...i_p} + \Gamma^k_{kj}
D^{ji_2...i_p} + (p-1)\,\Gamma^{i_2}_{kj}D^{kji_3...i_p}\ ,
\eeq
where the last term obviously vanishes. Therefore
\beq
{D^{kA_2...A_p}}_{|k} &=&  \partial_r D^{rA_2...A_p}
+ \partial_B D^{BA_2...A_p} \nonumber\\ & +&  \Gamma^r_{rr}
D^{rA_2...A_p} + \Gamma^{r}_{rB}D^{Bi_2...i_p} +
\Gamma^{B}_{Br}D^{rA_2...A_p} + \Gamma^{B}_{BC}D^{CA_2...A_p}\ .
\eeq
Now, taking into account (\ref{Gammas})
\beq
{D^{kA_2...A_p}}_{|k} &=& \left( \partial_r D^{rA_2...A_p}
+ \frac{2p}{r} D^{rA_2...A_p} \right)
+ \left( \partial_B D^{BA_2...A_p}
 + \Gamma^{B}_{BC}D^{CA_2...A_p} \right) \nonumber\\ &=&
r^{-2p}\, \partial_r (r^{2p}\,D^{rA_2...A_p}) +
{D^{A_1A_2...A_p}}_{||A_1}    \ .
\eeq

\hspace*{\fill}$\Box$

\section{Derivation of Maxwell equations}
\setcounter{equation}{0}
\label{MAX}

Using (\ref{dD}) one obtains:
\beq
\dot{Q}_{(1)}^{A_2...A_p} &=&  r \dot{D}^{rA_2...A_p} =
\frac{r}{p!} \epsilon^{rA_2...A_pA_1B_1...B_p}  B_{B_1...B_p||A_1}
\nonumber\\
&=& - \frac{r}{p!} \epsilon^{B_1...B_pA_1...A_p} B_{B_1...B_p||A_1} =
- \tr_{p-1}^{-1} \Pi^{A_2...A_p}_{(1)}\ .
\eeq
In the same way one proves (\ref{dQ2}) using (\ref{dB}). Now,
\beq
\dot{\Pi}^{(1)}_{B_2...B_p} &=& \frac{r}{p!} \tr_{p-1}^{-1}
\left( \epsilon_{A_1...A_pB_1...B_p} \dot{B}^{A_1...A_p||B_1} \right)
\nonumber\\
&=& \frac{r}{p!} \tr_{p-1}^{-1} \epsilon_{A_1...A_pB_1...B_p} \nabla^{B_1}
\left( (-1)^p\frac{1}{p!} \epsilon^{A_1...A_pkj_1...j_p}\nabla_k D_{j_1...j_p}
\right) \nonumber\\
&=& (-1)^p\frac{r}{(p!)^2} \tr_{p-1}^{-1} \epsilon_{A_1...A_pB_1...B_p}
\nabla^{B_1}
\left( \epsilon^{A_1...A_prC_1...C_p}\partial_r D_{C_1...C_p}
\right.  \nonumber\\ & & \left.
+\ p \epsilon^{A_1...A_pC_1rC_2...C_p}\nabla_{C_1} D_{rC_2...C_p} \right)
\nonumber\\
&=& (-1)^p\frac{r}{(p!)^2} \tr_{p-1}^{-1} \delta^{\ C_1...C_p}_{[B_1...B_p]}
\nabla^{B_1} \left( (-1)^p p! \partial_r D_{C_1...C_p} + (-1)^{p+1}pp!
\nabla_{C_1} D_{rC_2...C_p} \right) \nonumber\\
&=&  \frac{r}{p!} \tr_{p-1}^{-1} \nabla^{B_1} \left[ p! \partial_r
D_{B_1...B_p} +  p(p-1)! \left( \nabla_{B_1} D_{rB_2...B_p}
- \nabla_{B_2} D_{rB_1B_3...B_p} \right. \right. \nonumber\\ & &
\left. \left. -\ \nabla_{B_p} D_{rB_2...B_{p-1}B_1}
\right)\right] \nonumber\\
&=& r \tr_{p-1}^{-1} \nabla^{B_1} \left[  \partial_r
D_{B_1...B_p}  - (-1)^p  \nabla_{[B_2} D^r_{\ B_3...B_p]B_1}\right]
\nonumber\\
&=& r \tr_{p-1}^{-1}  \left[  \partial_r
\nabla^{B_1} D_{B_1...B_p} - r^{-2}\tr_{p-1}D_{rB_2...B_p}
\right]\ .
\eeq
Now, for any tensor field $X_{A...}^{\ \ \ ...}$ one has:
\begin{equation}   \label{l}
r^2\nabla^A \left(\partial_r X_{A...}^{\ \ \ ...} \right) =
\partial_r \left( r^2\nabla^A X_{A...}^{\ \ \ ...} \right)\ .
\end{equation}
To prove (\ref{l}) note, that
$$   [\partial_r, \nabla_B ] = 0\ , $$
and, therefore
\begin{equation}
{\left( \partial_r X_{A...}^{\ \ \ ...} \right)}_{||B}\eta^{AB} =
\partial_r\left( \nabla_B X_{A...}^{\ \ \ ...} \right) \eta^{AB} =
\partial_r\left( \eta^{AB}\nabla_B X_{A...}^{\ \ \ ...} \right) -
\partial_r\left( \eta^{AB} \right) \nabla_B X_{A...}^{\ \ \ ...} \ .
\end{equation}
But
$$ \partial_r \eta^{AB} = -\frac{2}{r} \eta^{AB} \ , $$
and the formula (\ref{l}) follows. Using (\ref{l}) one obtains
\beq
\dot{\Pi}^{(1)}_{B_2...B_p} &=&
r \tr_{p-1}^{-1}  \left[ r^{-2} \partial_r \left( r^2
\nabla^{B_1}D_{B_1...B_p}\right) - r^{-2}\tr_{p-1}D_{rB_2...B_p} \right]\ .
\eeq
Now,
\beq
\partial_r \left(r^2 {D_{B_1...B_p}}^{||B_1} \right) &=& \partial_r \left(r^2
\eta_{B_2C_2}...\eta_{B_pC_p}
{D^{C_1...C_p}}_{||C_1} \right) \nonumber\\ &=&
r^{-2p+2} \eta_{B_2C_2}...\eta_{B_pC_p}
\partial_r \left(r^{2p} {D^{C_1...C_p}}_{||C_1} \right)\nonumber\\ & =&
 - r^{-2p+2} \eta_{B_2C_2}...\eta_{B_pC_p}
\partial_r^2 \left(r^{2p} {D^{rC_2...C_p}} \right) \ ,
\eeq
due to Lemma~\ref{GAUSS}. Finally,
\beq
\dot{\Pi}^{(1)}_{B_2...B_p} &=& - \eta_{B_2C_2}...\eta_{B_pC_p}
 \tr_{p-1}^{-1}
\left[  r^{-2p+1} \partial^2_r \left( r^{2p-1} Q_{(1)}^{C_2...C_p} \right)
 + r^{-2} \tr_{p-1} Q_{(1)}^{C_2...C_p} \right] \nonumber\\
&=& -  \tr_{p-1}^{-1}
\left[  r^{-1} \partial^2_r \left( r Q^{(1)}_{B_2...B_p} \right)
 + r^{-2} \tr_{p-1} Q^{(1)}_{B_2...B_p} \right]\ .
\eeq
The same arguments apply for $\dot{\Pi}^{(2)}$.
\hspace*{\fill}$\Box$


\end{document}